\documentclass[12pt]{article}

\usepackage{epsf}
\usepackage[dvips]{graphicx}
\usepackage{epic}
\usepackage{eepic}

\begin{document}

\def\bc{\begin{center}}
\def\ec{\end{center}}
\def\beq{\begin{equation}}
\def\eeq{\end{equation}}
\def\noi{\noindent}
\def\ol#1{\overline{#1}}

\def\at#1{\left. \right|^{}_{#1}}
\def\hs#1{\hspace*{#1cm}}
\def\la{\langle}
\def\ra{\rangle}

\def\av#1{\langle #1 \rangle}
\def\ave#1{\langle {#1} \rangle}
\def\avr#1#2{\langle {#1} \rangle^{}_{#2}}
\def\avA#1{\langle #1 \rangle_A^{}}
\def\avB#1{\langle #1 \rangle_B^{}}
\def\avAB#1{\avA {\avB {{#1}}}}
\def\avBA#1{\avB {\avA {{#1}}}}
\def\avt#1{\avA {\avB {\overline{#1}}}}

\def\pra#1{\prod_{{#1}=1}^A}
\def\prda#1{\prod_{{#1}=1}^A da_{#1}}
\def\prhda#1{\prod_{{#1}=1}^A \hat{d}a_{#1}}
\def\prTda#1{\prod_{{#1}=1}^A T_A(a_{#1}) da_{#1}}

\def\prb#1{\prod_{{#1}=1}^B}
\def\prdb#1{\prod_{{#1}=1}^B db_{#1}}
\def\prhdb#1{\prod_{{#1}=1}^B \hat{d}b_{#1}}
\def\prTdb#1{\prod_{{#1}=1}^B T_B(b_{#1}) db_{#1}}

\def\sumA#1{\sum_{{#1}=1}^A}
\def\sumB#1{\sum_{{#1}=1}^B}

\def\vybk{\{k_1,...,k_n\}}
\def\vybkd{\{k_{n+1},...,k_B\}}

\def\wtk{\widetilde{k}}
\def\olk{\overline{k}}
\def\whk{\widehat{k}}
\def\vybn{k'_1,...,k'_n}
\def\vybm{k''_1,...,k''_m}
\def\vybr{\olk_1,...,\olk_r}
\def\vybd{k_1,...,k_{B-n-m-r}}
\def\QQ{Q^{(12)}}
\def\ss{\sigma^{(j_1 j_2)}}
\def\ssot{\sigma^{(1 2)}}
\def\ssi{ g_B }
\def\sig#1{\sigma_{#1}}
\def\Xj#1{X_{j_{#1}}}
\def\Yj#1{Y_{j_{#1}}}
\def\pj#1{p_{j_{#1}}}
\def\qj#1{q_{j_{#1}}}
\def\Cai{\{a_i\}}
\def\Cbk{\{b_k\}}
\def\hda#1{\hat{d}a_{#1}}
\def\hdb#1{\hat{d}b_{#1}}

\def\sNN{\sigma_{\!N\!N}^{}}
\def\sNNsq{\sigma_{\!N\!N}^{2}}

\def\stot{{\sigma_{}^{tot}}}
\def\sinel{{\sigma_{}^{in}}}
\def\sel{{\sigma_{}^{el}}}

\def\stotkj{{\sigma_{kj}^{tot}}}
\def\sinelkj{{\sigma_{kj}^{in}}}
\def\selkj{{\sigma_{kj}^{el}}}

\def\stotk{{\sigma_{k}^{tot}}}
\def\sinelk{{\sigma_{k}^{in}}}
\def\selk{{\sigma_{k}^{el}}}

\def\stotNN{{\sigma_{NN}^{tot}}}
\def\sinelNN{{\sigma_{NN}^{in}}}
\def\selNN{{\sigma_{NN}^{el}}}

\def\stotAB{{\sigma_{AB}^{tot}(b)}}
\def\sndifAB{{\sigma_{AB}^{non-dif}(b)}}
\def\sdifAB{{\sigma_{AB}^{dif}(b)}}

\def\stotA{{\sigma_{pA}^{tot}(b)}}
\def\sndifA{{\sigma_{pA}^{non-dif}(b)}}
\def\sdifA{{\sigma_{pA}^{dif}(b)}}

\def\sprodA{{\sigma_{pA}^{prod}(b)}}
\def\snprodA{{\sigma_{pA}^{non-prod}(b)}}

\def\im{{\,\rm{Im}\,}}
\def\re{{\,\rm{Re}\,}}
\def\hmod#1{\left| #1 \right|^2_{}}
\def\mod#1{| #1 |^2_{}}

\def\prodkj{{\prod_{k,j}^{}}}
\def\prodk{{\prod_{k}^{}}}

\def\akj{{a_{kj}^{}}}
\def\akjs{{a_{kj}^{*}}}

\def\ak{{a_{k}^{}}}
\def\aks{{a_{k}^{*}}}

\def\oakj{{(1+i\akj)}}
\def\oakjs{{(1-i\akjs)}}

\def\ppbar{{$\textrm{p}\overline{\textrm{p}}$}\ }


\begin{center}
{\bfseries
Fluctuations of the number of participants and binary collisions \\
in AA-interactions at fixed centrality in Glauber approach}
\vskip 5mm

V. V. Vechernin$^{\dag}$ and H. S. Nguyen

\vskip 5mm

{\small
{\it
Department of High-Energy Physics, St.~Petersburg State University,\\
RU-198504 St.~Petersburg, Russia
}
\\
$\dag$ {\it
E-mail: vechernin@pobox.spbu.ru
}}
\end{center}

\vskip 5mm

\begin{abstract}

In the framework of classical Glauber approach
the analytical expressions for the variance of the number of
wounded nucleons and binary collisions
in AA interactions at given centrality
are presented.
Along with the optical approximation term
they contain the additional contact terms,
arising only in the case of nucleus-nucleus collisions.
The magnitude of the additional contributions,
e.g. for PbPb collisions at SPS energies,
at some values of the impact parameter
is larger
than the contribution of the optical approximation,
with their sum being
in a good agreement
with the results of independent Monte-Carlo simulations
of this process.
Due to these additional terms
the variance of the total number of participants
for peripheral PbPb collisions
and the variance of the number of collisions
at all values of the impact parameter
exceed several times the Poisson ones.
The correlator between the numbers of participants
in colliding nuclei at fixed centrality is also
analytically calculated.
\end{abstract}

\vskip 10mm

\section{Introduction}
\label{sec:Intr}

At present the considerable attention is devoted to
the experimental and theoretical investigations of
the multiplicity and transverse momentum fluctuations of charged particles
in high energy AA collisions (see
\cite{NA49fluct07}-\cite{Konchakovski06}
and references therein).
One expects the increase of the fluctuations
in the case of freeze-out close to
the QCD critical endpoint of the
quark-gluon plasma - hadronic matter
phase boundary line \cite{Stephanov99,MG06}.

The aim of the present paper is
to draw an attention
to another factor leading to the
increase of the fluctuations
in the case of AA interactions.
Namely the increase of the fluctuations
of the number of participants and binary collisions
due to multiple contact nucleon interactions in nucleus-nucleus collisions.

Clear that these fluctuations
lead to fluctuations in the number of
particle sources
and so directly impact on
the multiplicity and transverse momentum
fluctuations of produced charged particles
and also on the correlations
between them (see, for example, \cite{PLB00}-\cite{CERES08}).

In the present paper
the  analytical expressions
for the variance of the number of
wounded nucleons and binary collisions
in given centrality AA interactions
are obtained
taking into account the multiple contact NN interactions
(so-called loop contributions).
The calculations are fulfilled in the framework of classical
Glauber approach \cite{Bialas76},
having a simple probabilistic interpretation \cite{CzyzMaximon69,Formanek69}.
In contrast with purely Monte-Carlo simulations
the analytical calculations enable to understand
the origin of increased values of the fluctuations.

As a result we demonstrate that
the multiple contact NN interactions in AA scattering
lead in particular to the fact that,
e.g. for PbPb collisions at SPS energies,
the variance of the total number of participants
for peripheral collisions
and the variance of the number of collisions
at all values of the impact parameter
exceed a few times the Poisson ones.

The paper is organized as follows.
In section 2 in the framework of classical Glauber approach
we present the  analytical expression
for the variance of the number of
wounded nucleons in one of the colliding nucleus
at a fixed value of the impact parameter.
Along with the well known optical contribution
(which depends only on the total inelastic NN cross-section)
in the case of nucleus-nucleus collisions
there is the additional contact term,
depending on the profile of the NN interaction probability
in the impact parameter plane.

In section 3 we calculate the correlator between the numbers of participants
in colliding nuclei at fixed centrality
and as a consequence find the variance of the total
(in both nuclei) number of participants.

In section 4 in the framework of the same approach
we present
the  analytical expression for the variance of the number of
NN binary collisions in given centrality AA interactions.
Along with the optical approximation term
it also contains other terms,
which occur the dominant ones.
These terms also correspond to the multinucleon contact interactions and
arise only in the case of nucleus-nucleus collisions.

The derivations of all formulas
are taken into the appendices A, B and C.

All over the paper
the results of numerical calculations are presented
with the purpose to illustrate the obtained analytical results.
We control also the results of our analytical calculations
comparing them with the results obtained by purely Monte-Carlo simulations
of the nucleus-nucleus scattering.

Note that
we restrict our consideration by the region of
the impact parameter $\beta < R_A + R_B$, where the probability
of inelastic interaction
$\sigma^{}_{\!AB}(\beta)$
of two nuclei with radii $R_A$ and $R_B$
is close to unity.

\section{Variance of the participants number in one nucleus}
\label{sec:NAw}

At first we consider the variance $V[N^A_w(\beta)]$
of the number of participants $N^A_w(\beta)$ (wounded nucleons)
at a fixed value of the impact parameter $\beta$ in one
of the colliding nuclei $A$.
In the framework of pure classical, probabilistic approach
to nucleus-nucleus collisions,
formulated in \cite{Bialas76}, we find for
the mean value and for the variance of $N^A_w(\beta)$ the following expressions (see appendix~\ref{ap:A}):
\begin{equation}
\langle N^A_w(\beta) \rangle = A P(\beta) \ ,
\label{mean}
\end{equation}
\begin{equation}
V[N^A_w(\beta)] =AP(\beta)Q(\beta)+A(A-1)[Q^{(12)}(\beta)-Q^2(\beta)] \ ,
\label{disp}
\end{equation}
where $P(\beta)=1-Q(\beta)$. For $Q(\beta)$ and $Q^{(12)}(\beta)$ we have
(all integrations imply the integration over
two-dimensional vectors in the impact parameter plane):
\begin{equation}
Q(\beta)=\int da\ T_A(a) [1- f_B (a\!+\!\beta)]^B \ ,
\label{Q}
\end{equation}
\begin{equation}
Q^{(12)}(\beta)= \int da_1 da_2 T_A(a_1) T_A(a_2)
[1- f_B (a_1\!+\!\beta)- f_B (a_2\!+\!\beta)
+ g_B (a_1\!+\!\beta,a_2\!+\!\beta)]^B
\label{QQ}
\end{equation}
with
\begin{equation}
 f_B (a) \equiv \int db\ T_B(b) \sigma(a\!-\!b)\ ,
\label{fB}
\end{equation}
\begin{equation}
 g_B (a_1,a_2)\equiv \int db\ T_B(b)
 \sigma(a_{1}\!-\!b)\sigma(a_{2}\!-\!b) \ .
\label{gB}
\end{equation}
Here $T_A$ and $T_B$ are the profile functions of
the colliding nuclei $A$ and $B$.
The $\sigma (a)$ is the probability of inelastic interaction of two nucleons
at the impact parameter $a$.
We'll imply that $\sigma (a)$, $T_A$ and $T_B$ depend only
on the magnitude of their two-dimensional vector argument.
Hence
$f_B (a)=f_B (|a|)$ and $Q(\beta)=Q(|\beta|)$.

The formula (\ref{mean}) and the first term in formula (\ref{disp}) correspond to
the naive picture
(so-called optical approximation)
implying that in the case of AA-collision at the impact parameter $\beta$
one can use the binomial distribution for $N^A_w(\beta)$
(see, for example, \cite{Wong,Vogt}):
\beq
\wp_{opt}(N^A_w)=C^{N^A_w}_A \, P(\beta)^{N^A_w}\,Q(\beta)^{A-N^A_w},
\hs 1 P(\beta)=1-Q(\beta)
\label{opt}
\eeq
with some averaged probability $P(\beta)$ of inelastic interaction
of a nucleon of the nucleus $A$ with nucleons of the nucleus $B$.
At that the $P(\beta)$ is considered to be the same for all
nucleons of the nucleus $A$.
In the optical approximation one has
\beq
\av{N^A_w(\beta)}_{opt} =A P(\beta)\ , \hs 1
V[ N^A_w(\beta)]_{opt} =AP(\beta)Q(\beta) \ .
\label{opt:avr_var}
\eeq

The whole expression (\ref{disp}) for the variance
is the result of more accurate calculation
(see appendix~\ref{ap:A}),
when at first one calculates the probabilities of all binary NN-interactions,
taking into account
the impact parameter plane positions
of nucleons in the nuclei $A$ and $B$
and only then averages over nucleon positions:
\begin{equation}
V[N^A_w(\beta)] =\av{{N_w^A(\beta)}^2}-\av{N^A_w(\beta)}^2 \ ,
\label{defvar}
\end{equation}
where
\begin{equation}
\av X \equiv \avt X \equiv
\int \ol{X} \prTdb{k} \prTda{j} \ .
\label{defavr}
\end{equation}
Here $\ol{X}$ is the average value of some variate $X$ at fixed positions of
all nucleons in the nuclei $A$ and $B$; $\avr{\ }{A}$ and $\avr{\ }{B}$
denote averaging over positions of these nucleons with corresponding nuclear
profile functions.

In the limit $r_N\ll R_A, R_B$ the formulae (\ref{fB}) and (\ref{gB}) reduce to
\beq
 f_B (a) \approx  \sNN \,T_B(a) \ , \hspace*{10mm}
\ssi(a_1,a_2) \approx I(a_{1}-a_{2}) \cdot T_B((a_{1}+a_{2})/2)
\label{ssiAPPROX}
\eeq
with
\beq
\sigma_{\!N\!N}^{} \equiv  \int\!db\, \sigma (b)     \ , \hspace*{25mm}
I(a)\equiv \int db \, \sigma(b)\, \sigma(b+a) \ .
\label{I}
\eeq
Note that in this limit
the $Q(\beta)$ and hence
the mean value (\ref{mean}) and
the first term of the variance (\ref{disp})
depend only on the integral inelastic $N\!N$
cross-section $\sigma_{\!N\!N}^{}$,
but the $Q^{(12)}(\beta)$ entering the second term of
the variance (\ref{disp})
depends also on the shape of the function $\sigma (b)$
through the integral $I(a)$ (\ref{I}).

Note also that using of the simple approximation with the $\delta$-function:
$\sigma (b)=\sNN \delta(b)$
for $N\!N$ interaction gives the same result
(as going to the limit $r_N\ll R_A, R_B$)
only for the optical part of the answer,
which is expressed through $Q(\beta)$.
If someone tries to use the approximation $\sigma (b)=\sNN \delta(b)$
to calculate $\QQ(\beta)$,
he will get $I(a)=\sigma_{\!N\!N}^2 \delta(a)$
and $\ssi=\sigma_{\!N\!N}^2 \delta(a_{1}-a_{2}) \cdot T_B(a_{1})$,
what leads to infinite $\QQ(\beta)$ at $B\geq 2$.
Meanwhile, for any correct approximation of $\sigma (b)$
with $\sigma (b)\leq 1$ (in correspondence with its probabilistic
interpretation in
classical Glauber approach)
we find a finite answer for $\QQ(\beta)$.

In the quantum case in Glauber approximation
due to unitarity one has
\beq
\label{inb}
\sigma (b)\equiv \sinel(b)=\stot(b)-\sel(b)=2\im \gamma(b)-\mod{\gamma(b)}\geq 0   \ ,
\eeq
where the $\gamma(b)$ is the amplitude of $N\!N$ elastic scattering.
This leads to the restrictions:
$0\leq\stot (b)\leq 4$,\ \ $0\leq\sel (b)\leq 4$ and $0\leq\sinel (b)\leq 1$.
So in the quantum case the $\sigma (b)$
also admits a probabilistic interpretation \cite{CzyzMaximon69,Formanek69}.

In our numerical calculations we have used for $\sigma (b)$
the "black disk" approximation:
\beq
\sigma (b)= \theta(r_N-|b|)\ ,
\label{black}
\eeq
and Gauss approximation:
\beq
\sigma (b)=\exp(-b^2/r^2_N)\ .
\label{Gauss}
\eeq
In both cases $\sNN=\pi r^2_N$.
For the nuclear profile functions $T_A$ and $T_B$
we have used the standard Woods-Saxon approximation:
\beq
T_A(a)=\int\!dz\, \rho(r) \ , \hspace*{10mm}
r^2=a^2+z^2\ , \hspace*{10mm}
\rho(r)=\rho_0 \left( 1+\exp\frac{r-R_A}{\kappa} \right)^{-1}
\label{WS}
\eeq
with $R_A=R_0 A^{\frac{1}{3}}$,  $R_0$=1.07 fm, $\kappa$=0.545 fm and
$\rho_0$ fixed by the condition $\int\! da\, T_A(a)\!=\!1$.

\begin{figure}[t]
\centerline{
\includegraphics[width=90mm,angle=-90]{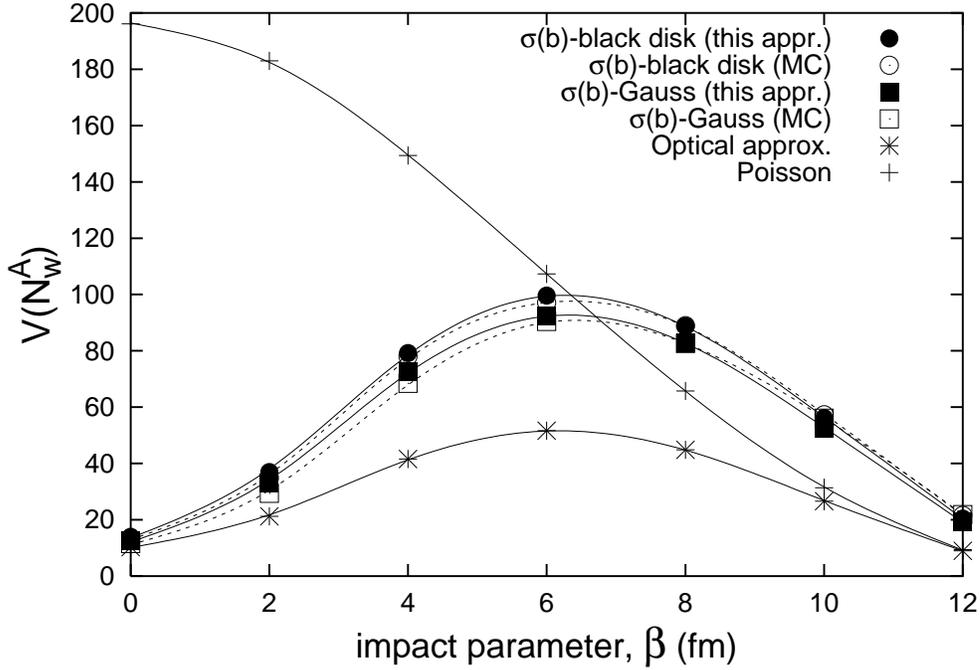}
}
\caption[dummy]{\label{vnwpoia}
The variance of the number of wounded nucleons in one nucleus
for PbPb collisions at SPS energies
($\sigma_{\!N\!N}^{}$=31\,mb)
as a function of the impact parameter $\beta$ (fm).\
The points {\Large $\bullet$} and $\rule[0.2mm]{2.3mm}{2.3mm}$
- results of numerical calculations
by the analytical formulae (\ref{disp})--(\ref{QQ}),
(\ref{ssiAPPROX}) and (\ref{I})
using respectively the black disk (\ref{black})
and Gaussian (\ref{Gauss}) approximations for $N\!N$ interaction; \
{\Large $\circ$} and \raisebox{1.5mm}{\framebox[2mm][c]{}}
 - results of independent MC simulations
using for $N\!N$ interaction
the black disk (\ref{black}) or Gaussian (\ref{Gauss}) approximation; \
{\Large \raisebox{0mm}{${\ast}$}}
- the optical approximation result~(\ref{opt:avr_var})
(the first term in formula (\ref{disp})); \
\raisebox{0mm}{+}
 - the Poisson variance: $V[N^A_w(\beta)]=\av{N^A_w(\beta)}$.
The curves are shown to guide eyes.
}
\end{figure}

The numerical evaluation of the contribution of
the additional (contact) term in formula (\ref{disp}) one can see
in Fig.\ref{vnwpoia} presented as an example for PbPb collisions at SPS energies
($r_N$=1\,fm, $\sigma_{\!N\!N}^{}$=31\,mb).
For the control we have also carried out independent
calculations of the mean values and the variances involved
by MC simulations of the AA scattering
presenting the results on the same figures.

In Fig.\ref{vnwpoia} we see that
the contact term in (\ref{disp})
is essential and
gives approximately the same contribution
to the variance of the $N^A_w(\beta)$ in PbPb collisions
at intermediate and large values of $\beta$ as the first optical term.
It's important that as we see in Fig.\ref{vnwpoia}
the results of independent MC simulations of the $N^A_w(\beta)$ variance
are in a good agreement with the results of the analytical calculations
by formula (\ref{disp})
only if one takes into account the contact term.

We see also in Fig.\ref{vnwpoia} that for peripheral AA collisions
at large $\beta$,
when $P(\beta)$ becomes small ($P(\beta)\ll$1, $Q(\beta)\approx 1$),
the optical approximation (\ref{opt})
reduces to the Poisson distribution
with $V[ N^A_w(\beta)]_{opt} \approx \av {N^A_w(\beta)}$ (\ref{opt:avr_var}).

So only due to the contact term the variance of the $N^A_w(\beta)$
is larger than the Poisson one
for peripheral PbPb collisions (at $\beta>7$\,fm)
in a correspondence with the indications,
which one has from the experimental data
on the dependence of multiplicity fluctuations on the
centrality at SPS and RHIC energies \cite{NA49fluct07,RHICpoisson}.

The week dependence of the results on the form
of $N\!N$ interaction at nucleon distances is also seen.
In the case of using the black disk (\ref{black}) approximation
for $\sigma (b)$ the results
lay systematically slightly higher,
than in the case of using the Gaussian (\ref{Gauss}) approximation
with the same value of $\sNN$.

\begin{figure}[t]
\centerline{
\includegraphics[width=90mm,angle=-90]{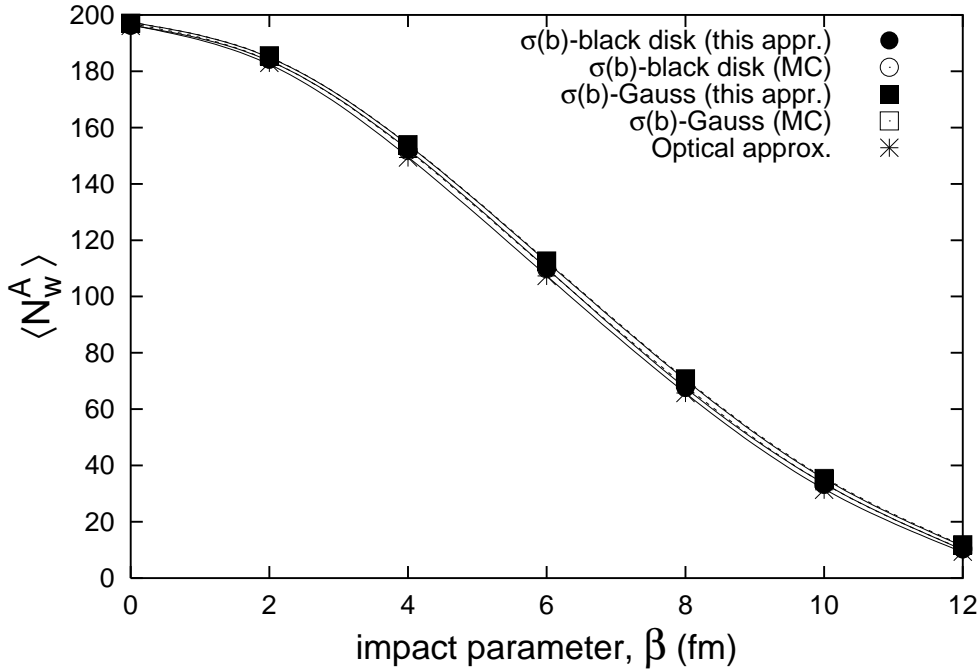}
}
\caption[dummy]{\label{nwa}
The same as in Fig.\ref{vnwpoia}, but for
the mean number of wounded nucleons in one nucleus,
calculated by formulae (\ref{mean}), (\ref{Q}), (\ref{fB})
and by independent MC simulations;
{\Large \raisebox{0mm}{${\ast}$}}
- the optical approximation result, calculated using formulae
(\ref{mean}), (\ref{Q}) and (\ref{I}).
}
\end{figure}

In Fig.\ref{nwa} we see that the mean value $\av{N^A_w(\beta)}$ (\ref{mean}),
in contrast to the variance,
coincides with the optical approximation result
(\ref{opt:avr_var})
and depends only on $\sNN$ in the limit $r_N\ll R_A, R_B$.
The MC simulations also confirm this result.

We would like to emphasize that the nontrivial term in
the expression (\ref{disp}) for the variance
arises only in the case of nucleus-nucleus collisions.
At $A=1$ or $B=1$ it vanishes.
At $A=1$ due to explicit factor $A-1$ in (\ref{disp})
and at $B=1$ due to fact that in this case $\QQ(\beta)=Q^2(\beta)$.
This corresponds to the well known fact that for nucleus-nucleus collisions
the Glauber approach doesn't reduce to the optical approximation
even in the limit $r_N\ll R_A, R_B$ (see, for example, \cite{Boreskov88}).

\begin{figure}[t]
\centerline{
\includegraphics[width=90mm,angle=0]{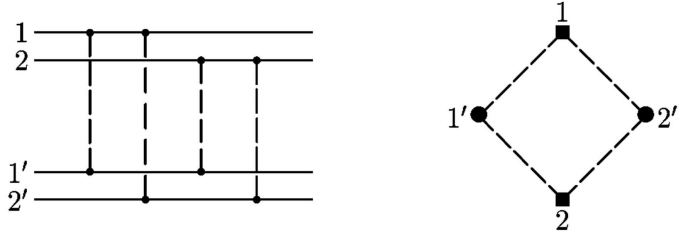}
}
\caption[dummy]{\label{loop}
An example of the loop diagram in AA-collisions.
1 and 2 - nucleons of the nucleus A;\ \
$1'$ and $2'$ - nucleons of the nucleus B
(see \cite{Boreskov88,Pak79,Braun878890} for details).
}
\end{figure}

The additional term,
which arises in the expression for the variance (\ref{disp}) in the case of nucleus-nucleus collisions,
depends, as we have mentioned, not only on the integral value
of inelastic $N\!N$ cross-section
$\sigma_{\!N\!N}^{}=\int\!db\, \sigma\! (b)$,
but also on the shape of the function $\sigma (b)$,
i.e. on the details of $N\!N$ interaction at
nucleon distances,
which are much smaller than the typical nuclear distances.
In quantum Glauber approach it corresponds to the fact that
in the case of AA collisions,
in contrast with pA collisions,
the loop diagrams of the type shown in Fig.\ref{loop} appear
and one encounters the contact terms problem
(see, for example, \cite{Boreskov88,Pak79,Braun878890}).

The second term in formula (\ref{disp}) is the manifestation of
this problem at the classical level.
In the case of a tree diagram the "lengths" of the interaction links
in the transverse plane
are independent. As a consequence the result expresses
only through $P(\beta)$ -
the probability  of the interaction
of a nucleon of the nucleus $A$ with nucleons of the nucleus $B$
{\it averaged} over its position in nucleus $A$.
The $P(\beta)$ is the same for any nucleon of the nucleus $A$.
In the case of the loop diagram in Fig.\ref{loop}
the "lengths" of the  interaction links in the transverse plane
are not independent and the result can't be expressed only through
the averaged probability $P(\beta)$
and the correlation effects have to be taken into account.

\section{Variance of the total number of participants}

Now we pass to the calculation of the variance of
the total number of participants
$V[N^A_w(\beta)+N^B_w(\beta)]$ at a fixed value
of the impact parameter $\beta$.
Clear, that for the mean value we have simply:
\beq
\av{N^A_w(\beta)+N^B_w(\beta)}= \av{N^A_w(\beta)} +\av{N^B_w(\beta)}
\label{meantot}
\eeq
and by (\ref{defvar}) for the variance
\beq
V[N^A_w(\beta)+N^B_w(\beta)] = V[N^A_w(\beta)]+V[N^B_w(\beta)]
+2\{ \av{N_w^A(\beta)N^B_w(\beta)}-\av{N^A_w(\beta)}\av{N^B_w(\beta)}\} \ .
\label{vartot}
\eeq

In naive optical approach there is no correlation
between the numbers of participants
in colliding nuclei
at fixed value of the impact parameter:
$$
\av{N_w^A(\beta)N^B_w(\beta)}_{opt}=\av{N^A_w(\beta)}_{opt}
\av{N^B_w(\beta)}_{opt}=
\av{N^A_w(\beta)} \av{N^B_w(\beta)} \ .
$$
More accurate calculations fulfilled in accordance with
(\ref{defvar}) and (\ref{defavr}) (see appendix~\ref{ap:B})
lead to
\begin{equation}
\av{N_w^A(\beta)N^B_w(\beta)}-\av{N^A_w(\beta)}\av{N^B_w(\beta)}=
AB[Q^{(11)}(\beta)-Q(\beta)\widetilde Q(\beta)] \ ,
\label{corr}
\end{equation}
where
\begin{equation}
Q^{(11)}(\beta)= \int da db T_A(a) T_B(b) [1- f_B (a\!+\!\beta)]^{B-1}
[1-f_A (b\!-\!\beta)]^{A-1}[1-\sigma(a\!-\!b\!+\!\beta)] \ ,
\label{Qoneone}
\end{equation}
\begin{equation}
\widetilde Q(\beta)=\int db\ T_B(b) [1- f_A (b\!-\!\beta)]^A
\label{tildeQ}
\end{equation}
and
\begin{equation}
f_A (b) \equiv \int da\ T_A(a) \sigma(b\!-\!a)
\approx \sNN \,T_A(b) \ .
\label{fA}
\end{equation}
The $Q(\beta)$ and $ f_B (a)$ are the same as in formulae
(\ref{Q}), (\ref{fB}) and (\ref{ssiAPPROX}).
Recall, that in our approximation
$f_A (b)=f_A (|b|)$ and $\widetilde Q(\beta)=\widetilde Q(|\beta|)$, then
$\widetilde Q(\beta)$ can be obtained from $Q(\beta)$
by a simple permutation
of $A$ and $B$. At $A=B$ we have $\widetilde Q(\beta)=Q(\beta)$.

\begin{figure}[t]
\centerline{
\includegraphics[width=90mm,angle=-90]{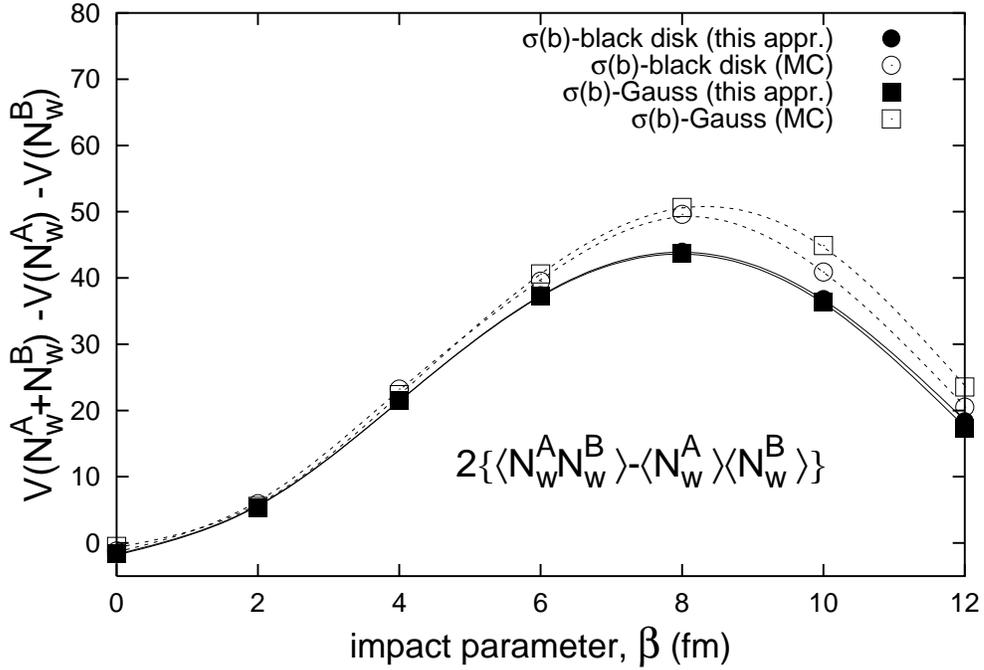}
}
\caption[dummy]{\label{vnw11ab}
The correlator between the numbers
of wounded nucleons in colliding nuclei,
calculated by analytical formulae (\ref{corr})-(\ref{fA})
and by independent MC simulations.
The notations are the same as in Fig.\ref{vnwpoia}.
}
\end{figure}

The results of numerical calculations of
the correlator (\ref{corr})
by formulae (\ref{Qoneone})--(\ref{fA})
for PbPb collisions at SPS energies
together with the results
obtained by independent MC simulations
of these collisions
are presented in Fig.\ref{vnw11ab}.

Comparing Fig.\ref{vnw11ab} with Fig.\ref{vnwpoia} we see that
the contribution of the correlator to the
variance of the total number of participants
at intermediate values of $\beta$
is about half of the variance for one nucleus $V[N^A_w(\beta)]$
and is approximately equal to the contribution of the first optical term in (\ref{disp}).
At large values of the impact parameter ($\beta\geq10$ fm)
the relative contribution of the correlator (\ref{corr})
to the total variance (\ref{vartot}) is even greater.
The results are again in a good agreement with the results
obtained by MC simulations.
(The small difference in the region 8-10 fm
arises from the use of approximate
formulae (\ref{ssiAPPROX}) and (\ref{fA}).)

\begin{figure}[t]
\centerline{
\includegraphics[width=90mm,angle=-90]{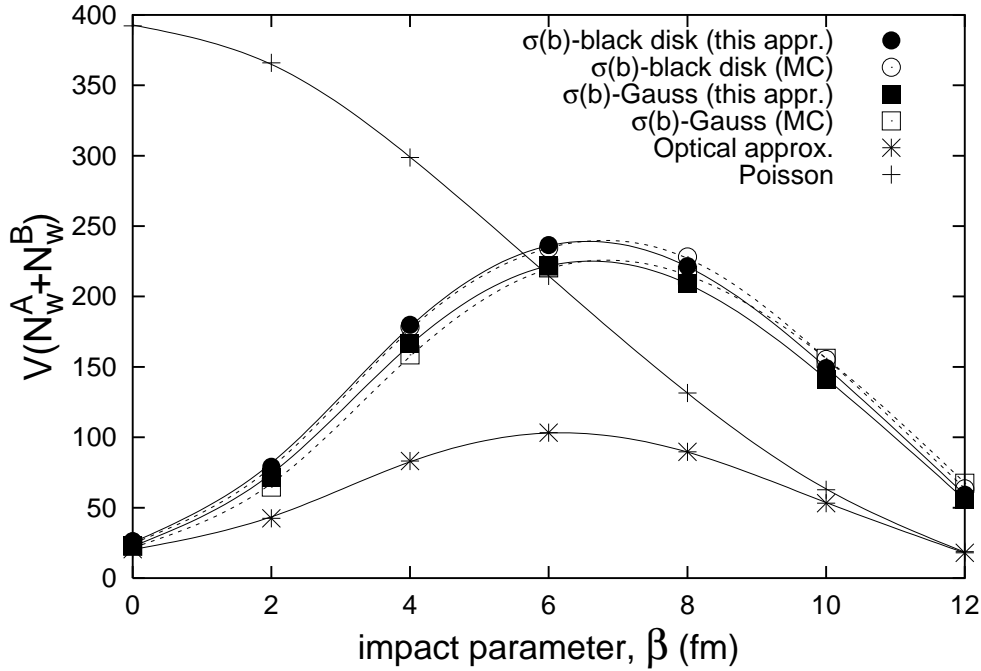}
}
\caption[dummy]{\label{vnwpoiab}
The same as in Fig.\ref{vnwpoia}, but for
the variance of the total number of wounded nucleons
$N_w(\beta)\equiv N^A_w(\beta)+N^B_w(\beta)$ in colliding nuclei.
The variance $V[N_w(\beta)]$ is calculated by formulae
(\ref{disp})--(\ref{QQ}), (\ref{ssiAPPROX}), (\ref{I})
with taking into account the contribution of
the correlator (\ref{vartot})-(\ref{fA});
\raisebox{0mm}{+}
 - the Poisson variance: \ $V[N_w(\beta)]=\av{N_w(\beta)}$.
}
\end{figure}

\begin{figure}
\centerline{
\includegraphics[width=90mm,angle=-90]{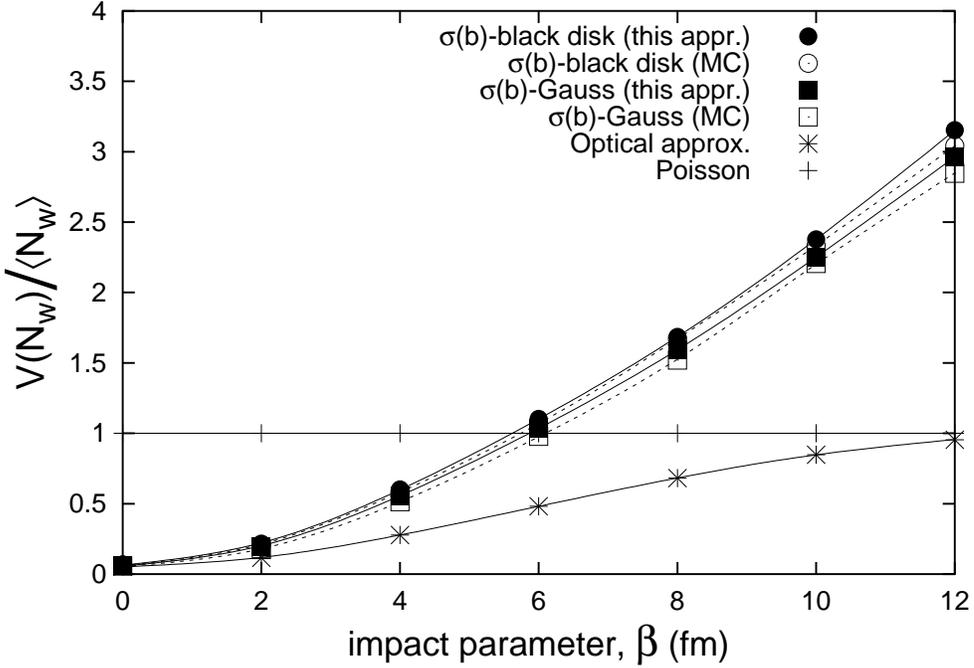}
}
\caption[dummy]{\label{vnw-nor}
The same as in Fig.\ref{vnwpoiab}, but for
the scaled variance
$V[ N_w(\beta)]/\av {N_w(\beta)}$
of the total number of wounded nucleons in colliding nuclei,
$N_w(\beta)\equiv N_w^A(\beta)+N^B_w(\beta)$.
}
\end{figure}

In Figs.\ref{vnwpoiab} and \ref{vnw-nor} we present the final results for the
variance of the total number of participants
in PbPb collisions at SPS energies,
taking into account the contribution of this correlator.
(In Fig.\ref{vnw-nor} the same, as in Fig.\ref{vnwpoiab},
but for the scaled variance:
$V[ N_w(\beta)]/\av {N_w(\beta)}$,
$N_w(\beta)\equiv N_w^A(\beta)+N^B_w(\beta)$.)
We see in particular that the calculated
variance of the total number of participants
$V[N^{}_w(\beta)]$
is a few times larger
than the Poisson one
in the impact parameter region 8-12\,fm.

\section{Variance of the number of binary collisions}

In this section we present the results of
the calculation of the variance of the number of NN-collisions
at a fixed value of the impact parameter $\beta$
in the framework of the same classical Glauber approach \cite{Bialas76}
to nucleus-nucleus collisions.
The details of calculations
one can find
in the appendix~\ref{ap:C}.

\begin{figure}[t]
\centerline{
\includegraphics[width=90mm,angle=-90]{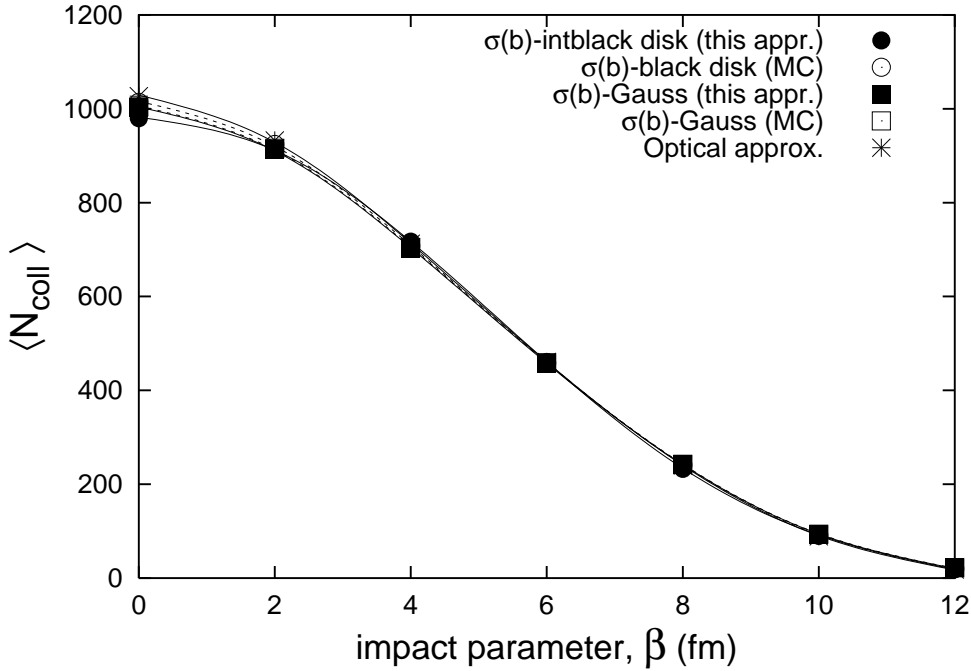}
}
\caption[dummy]{\label{ncoll}
The mean number of NN-collisions in PbPb interactions at SPS energies
calculated by the formulae (\ref{avNc}) and (\ref{chi})
and by independent MC simulations
as a function of the impact parameter $\beta$ (fm).
The notations are the same as in Fig.\ref{vnwpoia}.
}
\end{figure}

As a result we found that
the formula for the mean number of binary collisions
again coincides with the well-known expression given by
the optical approximation
(compare with the formula (\ref{opt:avr:coll}) below):
\begin{equation}
\av {N_{coll}(\beta)} = AB\chi(\beta) \ ,
\label{avNc}
\end{equation}
where
\begin{equation}
\chi(\beta) \equiv \int da db\ T_A(a) T_B(b) \sigma(a\!-\!b\!+\!\beta)
\approx \sNN \int da\ T_A(a) T_B(a\!+\!\beta)
\label{chi}
\end{equation}
has the meaning of the averaged probability of NN-interaction.
Numerically the mean value of the number of collisions as
a function of the impact parameter $\beta$ are shown in Fig.\ref{ncoll}.

In contrast to the mean value, the formula obtained
for the variance of $N_{coll}(\beta)$:
\begin{equation}
V[N_{coll}(\beta)] = AB[\chi(\beta)+(B\!-\!1)\chi^{}_1(\beta)
+(A\!-\!1)\widetilde \chi^{}_1(\beta)-(A\!+B\!-\!1) \chi^{2}(\beta)]
\label{VNc}
\end{equation}
differs from the optical approximation result (see below eq. (\ref{opt:var:coll})).
It depends not only on the $\chi(\beta)$ (\ref{chi}), but also on
\begin{equation}
\chi^{}_1(\beta) \equiv \int da \ T_A(a)
\left[\int db\ T_B(b) \sigma(a\!-\!b\!+\!\beta)\right]^2
\approx \sNNsq \int da\ T_A(a) T_B^2(a\!+\!\beta)
\label{chione}
\end{equation}
and
\begin{equation}
\widetilde \chi^{}_1(\beta) \equiv \int db\ T_B(b)
\left[\int da \ T_A(a) \sigma(a\!-\!b\!+\!\beta)\right]^2
\approx \sNNsq \int da\ T_B(a) T_A^2(a\!+\!\beta) \ .
\label{tchione}
\end{equation}
The $\widetilde \chi^{}_1$ is obtained from $\chi^{}_1$ by permutation
of $A$ and $B$. (Recall, that we consider
the $T_A$ and $T_B$ depend only
on the magnitude of their two-dimensional vector argument.)
At $A=B$ we have $\widetilde \chi^{}_1=\chi^{}_1$.
Note also that in the limit
$r_N\!\ll\!R_A, R_B$
the $\chi$, $\chi^{}_1$, $\widetilde \chi^{}_1$
and hence  the variance (\ref{VNc})
depend only on $\sNN$, but not on the
form of the function $\sigma(b)$
(it was not the case for the variance of the number of the wounded nucleons,
see section \ref{sec:NAw} after the formula (\ref{I})).

For comparison we list below the optical approximation results,
which assumes
the binomial distribution
for $N_{coll}(\beta)$ with the averaged probability $\chi(\beta)$
of NN-interaction
(see, for example, \cite{Wong,Vogt}):
\beq
\wp_{opt}(N_{coll})=C^{N_{coll}}_{AB}
\ \chi(\beta)^{N_{coll}}\,[1-\chi(\beta)]^{AB-N_{coll}}\ .
\label{opt:coll}
\eeq
In this case one has
\beq
\av{N_{coll}(\beta)}_{opt} =AB \chi(\beta)
\label{opt:avr:coll}
\eeq
and
\beq
V[ N_{coll}(\beta)]_{opt} =AB\chi(\beta)[1-\chi(\beta)] =
\av {N_{coll}(\beta)}[1-\chi(\beta)] \ .
\label{opt:var:coll}
\eeq
Note that for heavy nuclei
$\chi(\beta)$ is small even for central collisions
($\chi(\beta)\sim r^2_N/R^2_A \ll $1),
so the distribution (\ref{opt:coll}) and the variance
in optical approximation (\ref{opt:var:coll})
practically coincide with the Poisson ones:
$V[ N_{coll}(\beta)]_{opt} \approx \av {N_{coll}(\beta)}$.

Note also that in the case of pA interactions ($A=1$ or $B=1$)
our result (\ref{VNc}) for the variance of the number of collisions
coincides with the formula (\ref{opt:var:coll}) obtained
in the optical approximation.

\begin{figure}[t]
\centerline{
\includegraphics[width=90mm,angle=-90]{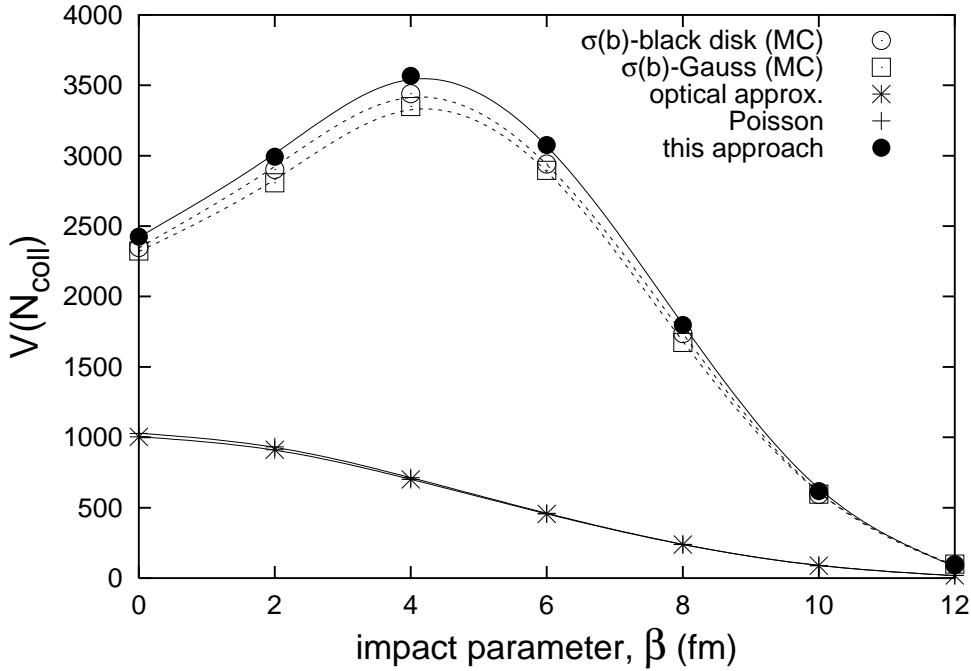}
}
\caption[dummy]{\label{vncoll}
The variance of the number of NN-collisions in PbPb interactions
at SPS energies as a function of the impact parameter $\beta$ (fm).
The points {\Large $\bullet$}
- results of calculations
by analytical formulae (\ref{chi})--(\ref{tchione}); \ \
{\Large \raisebox{0mm}{${\ast}$}}
- the optical approximation result, calculated using formulae
(\ref{chi}) and (\ref{opt:var:coll}); \ \
\raisebox{0mm}{+}
 - the Poisson variance: $V[N_{coll}(\beta)]=\av{N_{coll}(\beta)}$.
The notations are the same as in Fig.\ref{vnwpoia}.
}
\end{figure}

\begin{figure}[t]
\centerline{
\includegraphics[width=90mm,angle=-90]{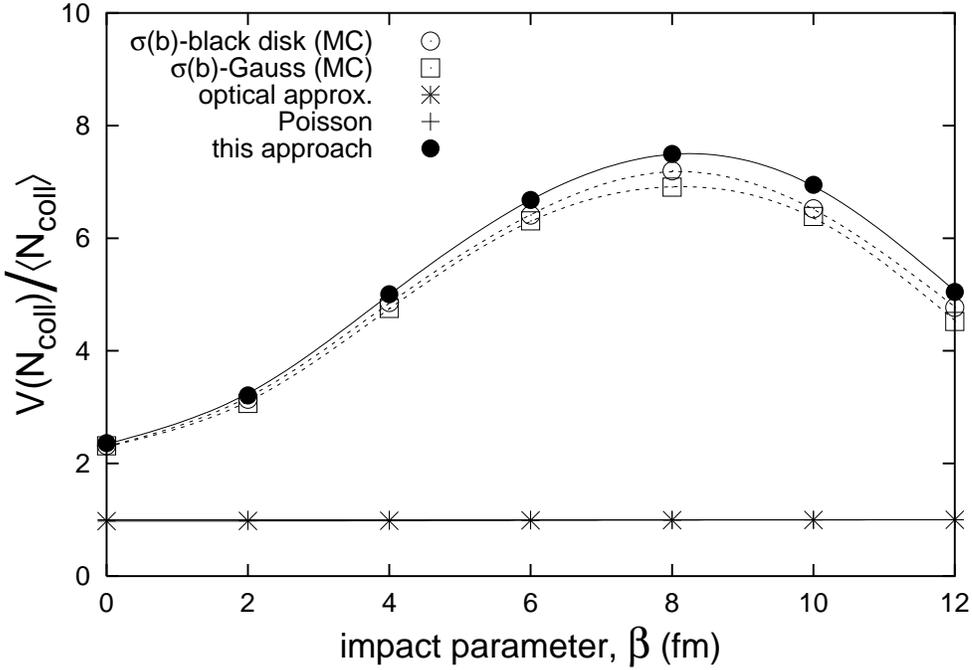}
}
\caption[dummy]{\label{vncl-nor}
The same as in Fig.\ref{vncoll}, but for
the scaled variance $V[ N_{coll}(\beta)]/\av {N_{coll}(\beta)}$
of the number of NN-collisions.
}
\end{figure}

 In Figs.\ref{vncoll} and \ref{vncl-nor},
 as an illustration we present,
the results of our numerical calculations of
the variance of the number of collisions
by analytical formulae (\ref{chi})--(\ref{tchione})
in the case of PbPb scattering at SPS energies
together with the results obtained from our independent
Monte-Carlo simulations
of the scattering process.
(In Fig.\ref{vncl-nor} the same as in Fig.\ref{vncoll}, but for
the scaled variance: $V[ N_{coll}(\beta)]/\av {N_{coll}(\beta)}$.)

We see that the calculated
variance of the number of collisions
at all values of the impact parameter $\beta$
is a few times larger than the Poisson one,
whereas the variance given
by the optical approximation practically
coincide with the Poisson one
(see the remark after formula (\ref{opt:var:coll})).
The results obtained by independent Monte-Carlo simulations
confirm our analytical result.
(The small difference again can be explained by the use of approximate
formulae (\ref{chi}), (\ref{chione})  and (\ref{tchione}).)

\begin{figure}[t]
\centerline{
\includegraphics[width=90mm,angle=-90]{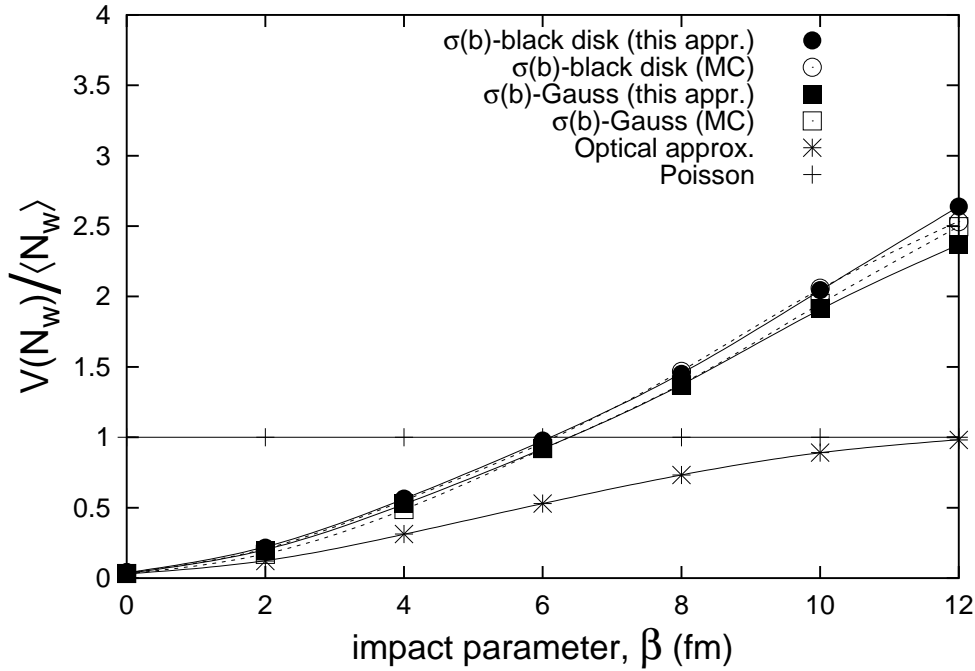}
}
\caption[dummy]{\label{vnw-nor-}
The scaled variance of the total number of wounded nucleons.
The same as in Fig.\ref{vnw-nor}, but for
the nucleon density distribution in nuclei  (\ref{WS}) with
a smaller value of the Woods-Saxon parameter $\kappa$=0.3 fm.
}
\end{figure}

\begin{figure}[t]
\centerline{
\includegraphics[width=90mm,angle=-90]{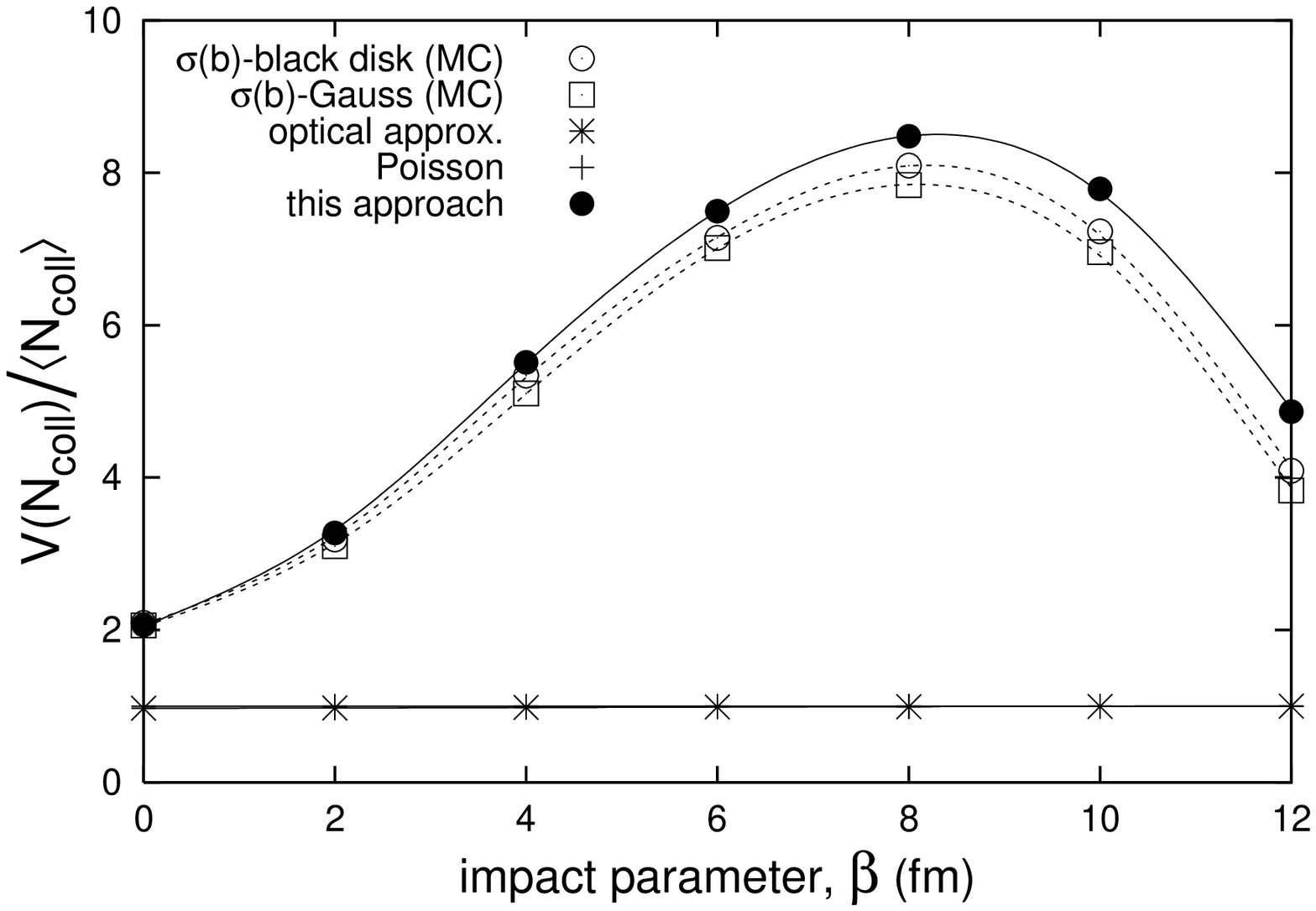}
}
\caption[dummy]{\label{vncl-nor-}
The scaled variance of the number of binary NN-collisions.
The same as in Fig.\ref{vncl-nor}, but for
the nucleon density distribution in nuclei (\ref{WS}) with
a smaller value of the Woods-Saxon  parameter $\kappa$=0.3 fm.
}
\end{figure}

We have also analyzed the dependence of the fluctuations on the
diffuseness of the nucleon density distribution in nuclei.
 To study this dependence the calculations
 with a smaller (0.3 fm) than standard (0.545 fm) value
of the  Woods-Saxon parameter $\kappa$  (\ref{WS}) were performed,
what corresponds to the model of nucleus
with a sharper edge  (see Figs.\ref{vnw-nor-} and \ref{vncl-nor-}).

The calculations confirm that one would expect from simple physical considerations,
more compact  distribution of  nucleons in nuclei
does not change the mean number of wounded nucleons,
but reduces its fluctuations, because in this case
the number of wounded nucleons is  more strictly determined
by the collision geometry.   As a result, the scaled variance
 of the number of wounded nucleons decrease with $\kappa$
 (compare the Figs.\ref{vnw-nor} and \ref{vnw-nor-}).

As for the number of binary NN-collisions,
in this case due to more compact  distribution of nucleons in nuclei
the mean number of collisions increases along with its variance.
Therefore the scaled variance
of the number of binary collisions
weakly depends on the variation
of the parameter $\kappa$
 (compare the Figs.\ref{vncl-nor} and Fig.\ref{vncl-nor-}).
 Important that in both cases the contribution of the contact term
 plays the crucial role.

\section{Discussion and conclusions}

It's shown that although the so-called optical approximation
gives the correct
results for the average number of wounded nucleons and binary collisions
the corresponding variances can't be described within this approximation
in the case of nucleus-nucleus interactions.

In the framework of classical Glauber approach
the  analytical expression for the variance of the number of participants
(wounded nucleons) in AA collisions at a fixed value of the impact parameter
is presented.
It's shown, that along with the optical approximation contribution
depending only on the total inelastic NN cross-section,
in the case of nucleus-nucleus collisions
there is the additional contact term contribution,
depending on details
of NN interaction at nucleon distances.

In classical Glauber approach
this contact contribution arises
due to taking into account
the interactions between two pairs of nucleons in colliding nuclei
(a pair in one nucleus with a pair in another).
It's found, that the interactions of higher order,
than between two pairs of nucleons,
don't contribute to the variance.
Whereas the expression for the mean number of participants
was proved to be exact already in the optical approximation,
which bases on taking into account only the averaged probability
of interaction between single nucleons in projectile and target nuclei.

These results are obtained in
the framework of
pure classical (probabilistic) Glauber approach \cite{Bialas76}.
However it's possible to suppose,
that in the quantum case
the one-loop expression for the variance
and the "tree" expression for the mean number of participants
and binary collisions will be exact.

Using obtained analytical formulae, the numerical calculation of
the variance of the participants number
in PbPb collisions at SPS energies was done as an example.
Demonstrated that
at intermediate and large impact parameter values
the optical and contact term contributions
are of the same order and
their sum is in a good agreement
with the results of independent MC simulations
of this process.

When calculating the variance of the total
(in both nuclei) number of participants
the correlation between the numbers of participants
in colliding nuclei is taking into account.
The  analytical expression for the correlator
at a fixed value of the impact parameter
is obtained.
The results of numerical calculations of
the correlator for the same process of PbPb collisions
show that
at intermediate and large values of
the impact parameter
its contribution to
the variance of the total number of participants
is about half of the variance in one nucleus,
again in good agreement with
independent MC simulations.

As a result
for peripheral PbPb collisions
the variance of the total number of participants,
calculated with taking into account
the contributions of this correlator and the contact terms,
occurs a few times larger than the Poisson one.

In the framework of the same classical Glauber approach
the analytical expression for the variance of the number of
NN binary collisions in given centrality AA interactions
is also found.
Along with the optical approximation term
it also contains other terms,
which occur the dominant ones.

Due to these additional terms
the variance of the number of collisions
at all values of the impact parameter
is several times higher than the Poisson one,
whereas the variance given
by the optical approximation practically
coincides with the Poisson one.
Again the results obtained by the independent MC simulations
confirm our analytical result.

Important that
these additional contact terms
in the expressions for the variances
arise only in the case of nucleus-nucleus collisions.
In the case of proton-nucleus collisions
they are missing
and the variances are well described
by the optical approximation.

Note that we have used  the simplest factorized approximation  (\ref{ap:factor})
for the nucleon density distribution in nuclei and
do not take into account nucleon-nucleon correlations within one
nucleus, which play a fundamental role, for example, in the description
of particle production in nuclear collisions outside the domain
kinematically available for a production from NN-scattering
(so-called 'cumulative' phenomena) \cite{FrStr8188}.

The additional contact contribution
to the variance of the number of wounded nucleons,
as we have found, arises
due to interactions between two pairs of nucleons in colliding nuclei,
which need to occur at the same position in the impact parameter plane.
Taking into account nucleon-nucleon correlations within one
nucleus must increase the probability
of such configurations
and hence
the contribution of the contact term.
However, numerical accounting of these effects is beyond the scope of the present paper.

Interestingly,  the nontrivial contact terms in variances
(missing in optical approximation)
arise in our approch already in the framework of the exploited factorized approximation
for the nucleon density in nuclei,
i. e. without taking into account nucleon-nucleon correlations within one
nucleus.

The authors thank M.A.~Braun and G.A.~Feofilov for useful discussions.
The work was supported by the RFFI grant 09-02-01327-a.

\section*{Appendices}
\appendix
\section{Calculation of the variance of participants in one nucleus}
\label{ap:A}

\begin{figure}[t]
\centerline{
\includegraphics[width=90mm,angle=0]{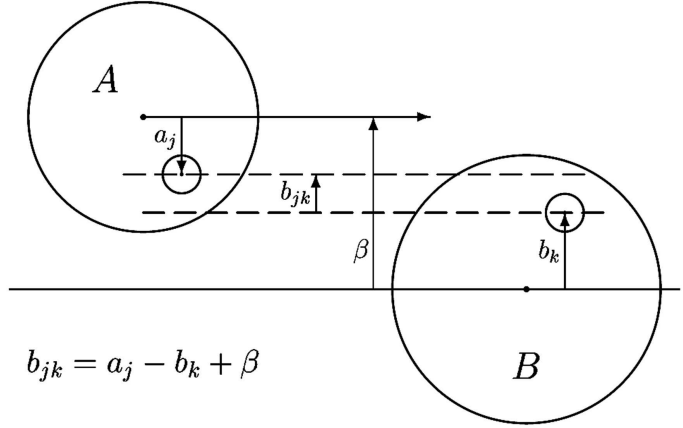}
}
\caption[dummy]{\label{AB}
Geometry of $AB$-collision.
}
\end{figure}

The geometry of $AB$-collision is depicted in Fig.\ref{AB}.
All $a_j$ and $b_k$ are the two-dimensional vectors
in the impact parameter plane.
In the framework of the classical (probabilistic) approach \cite{Bialas76}
the dimensionless $\sigma (b)$ is the probability
of inelastic interaction of two nucleons
at the impact parameter value~$b$ (see also (\ref{inb})).
The $T_A$ and $T_B$ are the profile functions
of the colliding nuclei $A$ and $B$.
We are implying that for heavy nuclei the factorization takes place:
\beq
T_A(a_1,...,a_A)=\prod_{j=1}^A T_A(a_j)  \ .
\label{ap:factor}
\eeq

Convenient to introduce the abbreviated notation:
\beq
\int\hat{d}a = \int T_A(a) \,da = 1 \ .
\label{short}
\eeq
All integrations imply the integration over
two-dimensional vectors in the impact parameter plane.
In new notation the (\ref{defavr}) takes the form
\beq
\av X \equiv \avt X = \int \ol{X} \prhdb{k} \prhda{j} \ .
\label{avr}
\eeq
Recall that here $\ol{X}$ means average of some variate $X$
at fixed positions of
all nucleons in $A$ and $B$; $\avr{\ }{A}$ and $\avr{\ }{B}$
mean averaging over positions of these nucleons.

We introduce the set of variates $X_1,...,X_A$ (each can be
equal only to 0 or 1) by the following way:
$X_j = 1$, if $j$-th nucleon of the nucleus $A$ interacts
with some nucleons of the nucleus $B$ and
$X_j = 0$, if $j$-th nucleon doesn't interact
with any nucleons of the nucleus~$B$.
The number of participants (wounded nucleons) in the nucleus $A$
in a given collision at the impact parameter $\beta$ is equal
to the sum of these variates:
\beq
N^A_w(\beta)=\sumA{j} X_j  \ .
\label{NAXj}
\eeq
Then we have for the mean value:
\beq
\av {N^A_w(\beta)} =\sumA{j} \av{X_j} =\sumA{j} \avt{X_j}
\label{avrNA}
\eeq
and for the variance of $N^A_w(\beta)$:
\beq
V[ N^A_w(\beta)] \equiv \av{{N_w^A(\beta)}^2}-\av{N^A_w(\beta)}^2 \ , \hs1
\av{{N_w^A(\beta)}^2}=\av{ (\sumA{j} X_j)^2 } \ .
\label{varNA}
\eeq

At first we calculate the mean value (\ref{avrNA}).
We denote by $q_j$ and $p_j$ the probabilities that the variate $X_j$
will be equal to 0 or 1 correspondingly.
Clear that for given configurations of nucleons $\{a_j\}$ and $\{b_k\}$ in nuclei $A$ and $B$:
\beq
q_j =  \prb{k}(1-\sig{jk}) \ , \hs1 p_j= 1-q_j \ ,
\label{PX1}
\eeq
where
\beq
\sig{jk}\equiv\sigma(a_j-b_k+\beta)
\label{sigjk}
\eeq
and
\beq
\ol{X_j}
= 0 \cdot q_j+1 \cdot p_j
= p_j \ .
\label{olXj}
\eeq
Note that $p_j$ and $q_j$ are the functions
of $a_j$, $b_1$,...,$b_B$ and $\beta$:
\beq
q_j=q_j(a_j,\Cbk,\beta)\ ,
\hs 1
p_j=p_j(a_j,\Cbk,\beta) \ .
\label{qjpj}
\eeq

Recall that
we restrict our consideration by the region of
the impact parameter $\beta < R_A + R_B$ where the probability
of inelastic nucleus-nucleus interaction
$\sigma^{}_{\!AB}(\beta)$
is close to unity.
Otherwise one has to introduce in formula (\ref{PX1}) for $q_j$
the factor $1/\sigma^{}_{\!AB}(\beta)$, where
\beq
\sigma^{}_{\!AB}(\beta) = 1-\avBA{\pra{j}\prb{k}(1-\sig{jk})}
\label{sigAB}
\eeq
and $\sigma^{}_{\!AB}=\int d\beta\, \sigma^{}_{\!AB}(\beta)$
is so-called production cross section, which
can't be calculated in a closed form.

Substituting (\ref{PX1})-(\ref{olXj}) into (\ref{avrNA})
we have
\beq
\av {N^A_w(\beta)}
=A-\sumA{j} \avAB{q_j} \ .
\label{ap:NAw}
\eeq
Averaging at first on positions of the nucleons in the nucleus $B$, we find
$$
\avB{q_j}
=(1-\sig{j})^B \ ,
$$
where we have introduced the short notation:
\beq
\sig{j} \equiv \int \hdb{1} \sig{j1} =
\int db_1 T_B(b_1) \sigma(a_j-b_1+\beta) \ .
\label{sigj}
\eeq
Averaging now on positions of the nucleons in the nucleus $A$, we have
\beq
\avAB{q_j}
=\int \hda{j} (1-\sig{j})^B  \ ,
\label{ap:qj}
\eeq
which is the same for any $j$, as the $a_j$ is the integration variable:
\beq
\avAB{q_j}=\int da_1 T_A(a_1) (1- \sig{1} )^B\equiv Q(\beta) \ .
\label{Q:ap}
\eeq
Then by (\ref{ap:NAw}) we find
\beq
\av {N^A_w(\beta)}
=A (1-Q(\beta))=A P(\beta) \ ,
\label{avNw}
\eeq
which coincides with formula (\ref{mean}) of the text,
if one takes into account the connection
\beq
\sig{j}=f_B(a_j+\beta)
\label{sigjf}
\eeq
(see (\ref{fB}) and (\ref{sigj})).
We see that the result for the mean number of participants (\ref{avNw})
is the same as in the optical approximation (\ref{opt:avr_var}).

We calculate now by the same way the variance of $N^A_w(\beta)$. By (\ref{varNA}) we have:
\beq
\av{{N_w^A(\beta)}^2}
=\sumA{j_1 \neq j_2} \av{ \Xj{1} \Xj{2} }+ \sumA{j} \av{X^2_j} \ .
\label{NA2}
\eeq
Note that the $\av{ \Xj{1} \Xj{2} }$ can't be reduced to the product
$\av{ \Xj{1}} \av{\Xj{2} }$.
Just in this point the optical approximation breaks for AB collisions.

Since  by (\ref{olXj})
$$
\ol{X^2_j} =\ol{X^{}_j} = p_j  \ ,
$$
then for the first sum in (\ref{NA2}) we find:
\beq
\sumA{j} \av{X^2_j}  =\sumA{j} \av{X^{}_j} =\av{{N_w^A(\beta)}}=AP(\beta) \ .
\label{S1}
\eeq
Because
$$
\ol{\Xj{1}\Xj{2}}=\ol{\Xj{1}}\cdot\ol{\Xj{2}}=\pj{1}\pj{2}
=1-\qj{1}-\qj{2}+\qj{1}\qj{2} \ ,
$$
for the second sum in (\ref{NA2}) using (\ref{Q:ap}) we have:
\beq
\sumA{j_1 \neq j_2} \av{ \Xj{1} \Xj{2} }
=A(A-1)[1-2Q(\beta)+\QQ(\beta)] \ ,
\label{S2}
\eeq
where we have introduced
\beq
\QQ(\beta)\equiv \frac{1}{A(A-1)}\sumA{j_1 \neq j_2} \avAB{ \qj{1}\qj{2} } \ .
\label{Q12}
\eeq

We calculate now $\QQ(\beta)$.
Averaging again at first on positions of the nucleons in the nucleus $B$,
we have
$$
\avB{ \qj{1}\qj{2} }
=(1-\sig{j_1}-\sig{j_2}+\ss)^B \ ,
$$
where $\sig{j_1}$ and $\sig{j_2}$ are given by (\ref{sigj}) and
\beq
\ss\equiv\int \hdb{1} \sig{j_1 1}\sig{j_2 1}
= \int db_1 T_B(b_1) \sigma(a_{j_1}-b_1+\beta)\sigma(a_{j_2}-b_1+\beta) \ .
\label{sigj12}
\eeq
Then averaging on positions of the nucleons in the nucleus $A$
one can rewright (\ref{Q12}) as follows
\beq
\QQ(\beta)
= \int da_1 da_2 T_A(a_1) T_A(a_2) (1- \sig{1}-\sig{2}+\ssot)^B \ ,
\label{ap:Q12}
\eeq
where by (\ref{sigj12})
\beq
\ssot=\int \hdb{1} \sig{1 1}\sig{2 1}
= \int db_1 T_B(b_1) \sigma(a_{1}\!-\!b_1\!+\!\beta)
\sigma(a_{2}\!-\!b_1\!+\!\beta)
\equiv \ssi (a_{1}\!+\!\beta, a_{2}\!+\!\beta) \
\label{sig12}
\eeq
(see notation (\ref{gB}) of the text).
Substituting (\ref{NA2}), (\ref{S1}) and (\ref{S2})
into (\ref{varNA}) we find
for the variance of $N^A_w(\beta)$:
$$
V[ N^A_w(\beta)]
=AQ(\beta)[1-Q(\beta)]+A(A\!-\!1)[\QQ(\beta)-Q^2(\beta)] \ ,
$$
which coincides with the formula (\ref{disp}) of the text
if we take into account (\ref{sigjf}), (\ref{ap:Q12}) and (\ref{sig12}).

\section{Correlation between the numbers
of participants in colliding nuclei at fixed centrality}
\label{ap:B}

The calculations are similar to ones in appendix~\ref{ap:A}
(we use the same notations).
Along with the set of variates $X_1,...,X_A$
we introduce in the symmetric way
the set of variates $\widetilde X_1,...,\widetilde X_B$
(each can be again equal only to 0 or 1).
$\widetilde X_k = 0(1)$ if $k$-th nucleon
of the nucleus $B$ doesn't interact (interacts)
with nucleons of the nucleus $A$.
Then similarly to (\ref{NAXj}) for the number
of participants (wounded nucleons)
in a given event in the nucleus $B$ we have:
\beq
N^B_w(\beta)=\sumB{k} \widetilde X_k  \ .
\label{NAtXk}
\eeq

Then
\beq
\av{N_w^A(\beta)N^B_w(\beta)}=
\sumA{j}\sumB{k} \avt{X_j \widetilde X_k}
\label{XjtXk}
\eeq
and  similarly to (\ref{olXj})
\beq
\ol{X_j \widetilde X_k}
= P_{jk}(1,1) \ ,
\label{Pjk}
\eeq
where the $P_{jk}(1,1)$
is the probability that the both variates $X_j$ and $\widetilde X_k$
will be equal to 1.
For the probability $P_{jk}(1,1)$ one finds
\beq
P_{jk}(1,1)=\sigma_{jk}+(1-\sigma_{jk})\rho_{jk}\widetilde\rho_{jk}  \ ,
\label{Pjk11}
\eeq
where $\sigma_{jk}$ is the probability of the interaction
of the $j$-th nucleon of the nucleus $A$
with the $k$-th nucleon of the nucleus $B$ (see formula (\ref{sigjk}))
and $\rho_{jk}$ is the probability of the interaction
of the $j$-th nucleon of the nucleus $A$
with at least one nucleon of the nucleus $B$ except the $k$-th nucleon
(correspondingly $\widetilde\rho_{jk}$ is the probability of the interaction
of the $k$-th nucleon of the nucleus $B$
with at least one nucleon of the nucleus $A$ except the $j$-th nucleon):
\beq
\rho_{jk}=1-\!\!\!\!\!\!\prod^B_{k'=1(k'\neq k)}
\!\!\!(1-\sigma_{jk'})\ , \hs 2
\widetilde\rho_{jk}=1-\!\!\!\!\!\!\prod^A_{j'=1(j'\neq j)}
\!\!\!(1-\sigma_{j'k})  \ .
\label{rhojk}
\eeq

Combining (\ref{XjtXk})--(\ref{rhojk}) and acting as in appendix~\ref{ap:A}
we find the formulae (\ref{corr})--(\ref{fA}) of the text.

\section{Fluctuations of the number of collisions}
\label{ap:C}

In this appendix we calculate the
variance of the number of NN-collisions
in AB-interaction at fixed value of centrality
in the framework of the approach under consideration.

To calculate the number of collisions we define the
set of the variates $Y_1,...,Y_A$, which can
take on a value from 0 to $B$.
If in the given event the $j$-th nucleon of the nucleus $A$ interacts
with $n$ nucleons of the nucleus $B$, then $Y_j = n $.
The number of NN-collisions in the given event at the impact parameter $\beta$
can be expressed through these variates as follows:
\begin{equation}
N_{coll}(\beta)=\sumA{j} Y_j
\label{coll}
\end{equation}
Clear that again (see appendix~\ref{ap:A}):
\begin{equation}
P(Y_j=0)= q_j =  \prb{k}(1-\sig{jk})
\label{PYj0}
\end{equation}
To calculate $P(Y_j=n)$ for $n=1,...,B$
we introduce $\vybk$ - the sampling from the set $\{1,...,B\}$
and $\vybkd$ - the rest after sampling. Then
\begin{equation}
P(Y_j=n) = \sum_{\vybk} \sig{jk_1}...\sig{jk_n}
(1-\sig{jk_{n+1}})...(1-\sig{jk_{B}})
\label{PYjn}
\end{equation}

First we again calculate the mean value of the number of collisions:
\begin{equation}
\av {N_{coll}(\beta)}
=\sumA{j} \avt{Y_j}  \ .
\label{Nc}
\end{equation}
For a given configuration $\{a_j\}$ and $\{b_k\}$ we have:
\begin{equation}
\ol{Y_j} =\sum_{n=0}^B \, n \, P(Y_j=n)  \ .
\label{olYj}
\end{equation}
Using (\ref{PYjn}) and averaging on positions of the nucleons in the nucleus $B$, one finds
\begin{equation}
\avB{\ol{Y_j}}
=\sum_{n=0}^B \, n \, C_B^n \sig{j}^n (1-\sig{j})^{B-n}
= B\,\sig{j}  \ .
\label{avYj}
\end{equation}
We use the same notations  as in appendix~\ref{ap:A} (see (\ref{sigj})).
Averaging then on positions of the nucleons in the nucleus $A$,
we finally find:
\begin{equation}
\av {N_{coll}(\beta)}= AB\chi(\beta) \ ,
\label{avNc:ap}
\end{equation}
where
\begin{equation}
\chi(\beta) \equiv \int \hda{1}  \sig{1}
=\int \hda{1} \hdb{1} \sig{11}
=\int da_1 db_1 T_A(a_1) T_B(b_1) \sigma(a_1-b_1+\beta) \ ,
\label{chi:ap}
\end{equation}
and at $r_N\ll R_A,R_B$
\begin{equation}
\chi(\beta) \approx \sigma_{\!N\!N}^{}\int da_1 T_A(a_1) T_B(a_1+\beta) \ ,
\label{approxchi}
\end{equation}
which coincides with the formulae (\ref{avNc}) and (\ref{chi}) of the text.
Comparing (\ref{avNc:ap}) and (\ref{opt:avr:coll})
we see that the result for the mean number of collisions
is the same as in the optical approximation.

In the rest of the appendix
we calculate the variance of the number of collisions.
To calculate the variance:
\beq
V[ N_{coll}(\beta)] \equiv \av{{N^2_{coll}(\beta)}}-\av{N_{coll}(\beta)}^2
\label{varNc}
\eeq
one has to calculate
\beq
\av{{N^2_{coll}(\beta)}}=\av{ (\sumA{j} Y_j)^2 }=
\sumA{j_1 \neq j_2} \av{ \Yj{1} \Yj{2} }+ \sumA{j} \av{Y^2_j} \ .
\label{Nc2}
\eeq
So we have to calculate the following two sums:
\beq
\sumA{j_1 \neq j_2} \av{ \Yj{1} \Yj{2} }=
\sumA{j_1 \neq j_2} \avt{ \Yj{1} \Yj{2} }
\label{Yj12}
\eeq
and
\beq
\sumA{j} \av{Y^2_j} = \sumA{j} \avt{Y^2_j} \ .
\label{Yj2}
\eeq

To calculate the first sum we
denote by $\vybn$ - the indices of the nucleons
of the nucleus $B$, which interact
only with the nucleon $j_1$ of the nucleus $A$.
By $\vybm$ we denote the indices of the nucleons,
which interact only with the nucleon $j_2$ of the nucleus $A$
and  by $\vybr$ we denote the indices of the nucleons,
which interact with both nucleons $j_1$ and $j_2$.
By $\vybd$ we denote the indices of the nucleons of the nucleus $B$,
which don't interact with the nucleons $j_1$ and $j_2$ of the nucleus $A$.
Then the probability $p_{j_1 j_2}$ of such event in these notations
is equal to
\begin{equation}
p_{j_1 j_2} = p_{j_1} p_{j_2} \ ,
\label{pj1j2}
\end{equation}
where
\begin{equation}
p_{j_1} =     \prod_{i=1}^r \sig{j_1\olk_i}
              \prod_{i=1}^n \sig{j_1k'_i}
              \prod_{i=1}^m (1-\sig{j_1k''_i})
              \prod_{i=1}^{B-r-m-n} (1-\sig{j_1k_i})  \ ,
\label{pj1}
\end{equation}
\begin{equation}
p_{j_2} =     \prod_{i=1}^r \sig{j_2\olk_i}
              \prod_{i=1}^n (1- \sig{j_2k'_i})
              \prod_{i=1}^m \sig{j_2k''_i}
              \prod_{i=1}^{B-r-m-n} (1-\sig{j_2k_i})  \ .
\label{pj2}
\end{equation}
Using (\ref{pj1}) and (\ref{pj2}) we can rewrite $p_{j_1 j_2}$
in the following form
\begin{equation}
p_{j_1 j_2} = \prod_{i=1}^r \sig{j_1\olk_i} \sig{j_2\olk_i}
              \prod_{i=1}^n \sig{j_1k'_i} (1- \sig{j_2k'_i})
              \prod_{i=1}^m (1-\sig{j_1k''_i})\sig{j_2k''_i}
         \!\! \prod_{i=1}^{B-r-m-n} \!\! (1-\sig{j_1k_i}-\sig{j_2k_i}
         +\sig{j_1k_i}\sig{j_2k_i}) \ .
\label{pj12}
\end{equation}

The probability $P_{j_1 j_2}(n,m,r)$ that the nucleons $j_1$ and $j_2$
of the nucleus $A$
interact separately with $n$ and $m$ nucleons
of the nucleus $B$
and at that else simultaneously with $r$ nucleons of the nucleus $B$
is equal to
\begin{equation}
P_{j_1 j_2}(n,m,r) = \sum p_{j_1 j_2}  \ ,
\label{Pnmr}
\end{equation}
where the sum means summing on all possible three sampling
$\{\vybn\}$, $\{\vybm\}$, $\{\vybr\}$ from the set $\{1,...,B\}$.
After averaging (\ref{Pnmr}) on positions of the nucleons in the nucleus $B$ we find
\begin{equation}
\avB{P_{j_1 j_2}(n,m,r)}=\frac{B!}{n!m!r!(B-r-m-n)!}
z^r (y-z)^m (x-z)^n (1-x-y+z)^{B-r-m-n}\ ,
\label{PnmrB}
\end{equation}
where we have used the short notations:
\beq
x=\sig{j_1}\ , \hs1
y=\sig{j_2}\ , \hs1
z=\ss\ . \hs1
\label{xyz}
\eeq
The $\sig{j_1}$ and $\sig{j_2}$ are defined by (\ref{sigj}) and
the $\ss$ is defined by (\ref{sigj12})
in appendix~\ref{ap:A}. Then for the components of the first sum (\ref{Yj12})
we have
\begin{equation}
\avr{ \ol{\Yj{1}\Yj{2}}  }B =\sum_{r=0}^B \sum_{m=0}^{B-r} \sum_{n=0}^{B-r-m}
(m\!+\!r)(n\!+\!r)\avB{P_{j_1 j_2}(n,m,r)}\ .
\label{avrY1Y2}
\end{equation}
After substitution of (\ref{PnmrB}) in (\ref{avrY1Y2})
the lengthy but straightforward calculation leads to the simple answer
\begin{equation}
\avr{ \ol{\Yj{1}\Yj{2}}  }B = Bz+B(B\!-\!1)xy =
B\ss+B(B\!-\!1)\sig{j_1}\sig{j_2} \ .
\label{avY1Y2}
\end{equation}
For the components of the second sum (\ref{Yj2}) the similar but much more simple
calculation gives
\begin{equation}
\avr{ \ol{\Yj{}^2}  }B = B\sig{j}+B(B\!-1\!)\sig{j}^2 \ .
\label{avY2}
\end{equation}

Averaging now on positions of the nucleons in the nucleus $A$,
we can rewrite (\ref{Nc2}) as
$$
\av{{N^2_{coll}(\beta)}}=
B\left[\sumA{j_1 \neq j_2} \left(\avr{\ss}A +
(B\!-\!1)\avr{\sig{j_1}\sig{j_2} }A\right)
+\sumA{j} \left( \avr{ \sig{j} }A+(B\!-\!1)\avr{\sig{j}^2 }A \right) \right]=
$$
$$
=B\left[A(A\!-\!1)\int \hda{1} \hda{2} \left(\avr{\ssot}A +
(B\!-\!1)\avr{\sig{1}\sig{2} }A\right)
+A \int \hda{1} \left( \avr{ \sig{1} }A+(B\!-\!1)\avr{\sig{1}^2 }A \right)
\right]
$$
Recalling now that $\sig{1}$, $\sig{2}$ and $\ssot$ are given
by the formulae (\ref{sigj}) and (\ref{sig12})
of the appendix~\ref{ap:A},
we obtain
\beq
\av{{N^2_{coll}(\beta)}}=AB[\chi(\beta)+(B\!-\!1)\chi^{}_1(\beta)
+(A\!-\!1)\widetilde \chi^{}_1(\beta)+(A\!-\!1)(B\!-\!1) \chi^{2}(\beta)]
\label{avNc2}
\end{equation}
with $\chi(\beta)$, $\chi^{}_1(\beta)$ and $\widetilde \chi^{}_1(\beta)$
defined by
the formulae (\ref{chi}), (\ref{chione}) and (\ref{tchione}) of the text.
Using now the definition (\ref{varNc})
and taking into account the formula (\ref{avNc:ap})
for $\av {N_{coll}(\beta)}$
we come to the expression (\ref{VNc}) of the text
for the variance of the number of collisions.

\end{document}